\documentclass[prd,aps,showpacs,epsf,floats,10pt]{revtex4}
\usepackage{amssymb}
\usepackage{amsfonts}
\usepackage{amsmath}

\input{tcilatex}
\begin{document}

\title{Works on an information geometrodynamical approach to chaos }
\author{Carlo Cafaro}
\email{carlocafaro2000@yahoo.it}
\affiliation{Department of Physics, State University of New York at Albany-SUNY,1400
Washington Avenue, Albany, NY 12222, USA}

\begin{abstract}
In this paper, I propose a theoretical information-geometric framework
suitable to characterize chaotic dynamical behavior of arbitrary complex
systems on curved statistical manifolds. Specifically, I present an
information-geometric analogue of the Zurek-Paz quantum chaos criterion of
linear entropy growth and an information-geometric characterization of
regular and chaotic quantum energy level statistics.
\end{abstract}

\maketitle


\section{\textbf{Introduction}}

The study of complexity \cite{gell-mann} has created a new set of ideas on
how very simple systems may give rise to very complex behaviors. Moreover,
in many cases, the "laws of complexity" have been found to hold universally,
caring not at all for the details of the system's constituents. Chaotic
behavior is a particular case of complex behavior and it will be the object
of the present work.

In this paper I make use of the so-called Entropic Dynamics (ED) \cite%
{caticha1}. ED is a theoretical framework that arises from the combination
of inductive inference (Maximum Entropy Methods (ME), \cite{caticha2}) and
Information Geometry (IG) \cite{amari}. The most intriguing question being
pursued in ED stems from the possibility of deriving dynamics from purely
entropic arguments. This is clearly valuable in circumstances where
microscopic dynamics may be too far removed from the phenomena of interest,
such as in complex biological or ecological systems, or where it may just be
unknown or perhaps even nonexistent, as in economics. It has already been
shown that entropic arguments do account for a substantial part of the
formalism of quantum mechanics, a theory that is presumably fundamental \cite%
{caticha3}. Perhaps the fundamental theories of physics are not so
fundamental; they may just be consistent, objective ways to manipulate
information. Following this line of thought, I extend the applicability of
ED to temporally-complex (chaotic) dynamical systems on curved statistical
manifolds and identify relevant measures of chaoticity of such an
information geometrodynamical approach to chaos (IGAC).

The layout of the paper is as follows. In the next Section, I give an
introduction to the main features of our IGAC. In Section III, I apply my
theoretical construct to three complex systems. First, I study the chaotic
behavior of an ED Gaussian model describing an arbitrary system of $l$
degrees of freedom and show that the hyperbolicity of the non-maximally
symmetric $2l$-dimensional statistical manifold $\mathcal{M}_{s}$ underlying
such ED Gaussian model leads to linear information geometrodynamical entropy
(IGE) growth and to exponential divergence of the Jacobi vector field
intensity. An information-geometric analogue of the Zurek-Paz quantum chaos
criterion of linear entropy growth and an information-geometric
characterization of regular and chaotic quantum energy level statistics are
presented.

I emphasize that I have omitted technical details that will appear
elsewhere. However, some applications of my IGAC to low dimensional chaotic
systems can be found in my previous articles \cite{cafaro1, cafaro2,
cafaro3, cafaro4}. Finally, in Section IV I present my conclusions and
suggest further research directions.

\section{The information geometrodynamical approach to chaos: General
Formalism}

The IGAC is an application of ED to complex systems of arbitrary nature. ED
is a form of information-constrained dynamics built on curved statistical
manifolds $\mathcal{M}_{S}$ where elements of the manifold are probability
distributions $\left\{ P\left( X|\Theta \right) \right\} $ that are in a
one-to-one relation with a suitable set of macroscopic statistical variables 
$\left\{ \Theta \right\} $ that provide a convenient parametrization of
points on $\mathcal{M}_{S}$. The set $\left\{ \Theta \right\} $ is called
the \textit{parameter space }$\mathcal{D}_{\Theta }$ of the system.

In what follows, I schematically outline the main points underlying the
construction of an arbitrary form of entropic dynamics. First, the
microstates of the system under investigation must be defined. For the sake
of simplicity, I assume the system is characterized by an $l$-dimensional 
\textit{microspace} with microstates $\left\{ x_{k}\right\} $ where $k=1$%
,..., $l$. These microstates can be interacting, non-interacting, or I may
have no relevant information concerning their microscopic dynamics. Indeed,
the main goal of an ED model is that of inferring "macroscopic predictions"
in the absence of detailed knowledge of the microscopic nature of arbitrary
complex systems. Once the microstates have been defined, I have to select
the relevant information about such microstates. In other words, I have to
select the \textit{macrospace} of the system. For the sake of the argument,
I assume that our microstates are Gaussian-distributed. They are defined by $%
2l$-information constraints, for example their expectation values $\mu _{k}$
and variances $\sigma _{k}$.%
\begin{equation}
\left\langle x_{k}\right\rangle \equiv \mu _{k}\text{ and }\left(
\left\langle \left( x_{k}-\left\langle x_{k}\right\rangle \right)
^{2}\right\rangle \right) ^{\frac{1}{2}}\equiv \sigma _{k}\text{.}
\end{equation}%
In addition to information constraints, each Gaussian distribution $%
p_{k}\left( x_{k}|\mu _{k}\text{, }\sigma _{k}\right) $ of each microstate $%
x_{k}$ must satisfy the usual normalization conditions,%
\begin{equation}
\dint\limits_{-\infty }^{+\infty }dx_{k}p_{k}\left( x_{k}|\mu _{k}\text{, }%
\sigma _{k}\right) =1\text{ }
\end{equation}%
where 
\begin{equation}
p_{k}\left( x_{k}|\mu _{k}\text{, }\sigma _{k}\right) =\left( 2\pi \sigma
_{k}^{2}\right) ^{-\frac{1}{2}}\exp \left( -\frac{\left( x_{k}-\mu
_{k}\right) ^{2}}{2\sigma _{k}^{2}}\right) \text{.}
\end{equation}%
Once the microstates have been defined and the relevant (linear or
nonlinear) information constraints selected, I am left with a set of
probability distributions $p\left( X|\Theta \right) =\underset{k=1}{\overset{%
l}{\dprod }}$ $p_{k}\left( x_{k}|\mu _{k}\text{, }\sigma _{k}\right) $
encoding the relevant available information about the system where $X$ is
the $l$-dimensional microscopic vector with components $\left( x_{1}\text{%
,...,}x_{l}\right) $ and $\Theta $ is the $2l$-dimensional macroscopic
vector with coordinates $\left( \mu _{1}\text{,..., }\mu _{l}\text{; }\sigma
_{1}\text{,..., }\sigma _{l}\right) $. The set $\left\{ \Theta \right\} $
define the $2l$-dimensional space of macrostates of the system, the
statistical manifold $\mathcal{M}_{S}$. A measure of distinguishability
among macrostates is obtained by assigning a probability distribution $%
P\left( X|\Theta \right) \ni \mathcal{M}_{S}$ to each macrostate $\Theta $ .
Assignment of a probability distribution to each state endows $\mathcal{M}%
_{S}$ with a metric structure. Specifically, the Fisher-Rao information
metric $g_{\mu \nu }\left( \Theta \right) $ \cite{amari}, 
\begin{equation}
g_{\mu \nu }\left( \Theta \right) =\int dXp\left( X|\Theta \right) \partial
_{\mu }\log p\left( X|\Theta \right) \partial _{\nu }\log p\left( X|\Theta
\right) \text{,}  \label{fisher-rao}
\end{equation}%
with $\mu $,$\nu =1$,..., $2l$ and $\partial _{\mu }=\frac{\partial }{%
\partial \Theta ^{\mu }}$ , defines a measure of distinguishability among
macrostates on $\mathcal{M}_{S}$. The statistical manifold $\mathcal{M}_{S}$,%
\begin{equation}
\mathcal{M}_{S}=\left\{ p\left( X|\Theta \right) =\underset{k=1}{\overset{l}{%
\dprod }}p_{k}\left( x_{k}|\mu _{k}\text{, }\sigma _{k}\right) \right\} 
\text{,}
\end{equation}%
is defined as the set of probabilities $\left\{ p\left( X|\Theta \right)
\right\} $ described above where $X\in 
\mathbb{R}
^{3N}$, $\Theta \in \mathcal{D}_{\Theta }=\left[ \mathcal{I}_{\mu }\times 
\mathcal{I}_{\sigma }\right] ^{3N}$. The parameter space $\mathcal{D}%
_{\Theta }$ (homeomorphic to $\mathcal{M}_{S}$) is the direct product of the
parameter subspaces $\mathcal{I}_{\mu }$ and $\mathcal{I}_{\sigma }$, where
(unless specified otherwise) $\mathcal{I}_{\mu }=\left( -\infty \text{, }%
+\infty \right) _{\mu }$ and $\mathcal{I}_{\sigma }=\left( 0\text{, }+\infty
\right) _{\sigma }$. Once $\mathcal{M}_{S}$ and $\mathcal{D}_{\Theta }$ are
defined, the ED formalism provides the tools to explore dynamics driven \ on 
$\mathcal{M}_{S}$\ by entropic arguments. Specifically, given a known
initial macrostate $\Theta ^{\left( \text{initial}\right) }$ (probability
distribution), and that the system evolves to a final known macrostate $%
\Theta ^{\left( \text{final}\right) }$, the possible trajectories of the
system are examined in the ED approach using ME methods.

I emphasize ED can be derived from a standard principle of least action (of
Jacobi type). The geodesic equations for the macrovariables of the Gaussian
ED model are given by\textit{\ nonlinear} second order coupled ordinary
differential equations,%
\begin{equation}
\frac{d^{2}\Theta ^{\mu }}{d\tau ^{2}}+\Gamma _{\nu \rho }^{\mu }\frac{%
d\Theta ^{\nu }}{d\tau }\frac{d\Theta ^{\rho }}{d\tau }=0\text{.}
\label{geodesic equations}
\end{equation}%
The geodesic equations in (\ref{geodesic equations}) describe a \textit{%
reversible} dynamics whose solution is the trajectory between an initial $%
\Theta ^{\left( \text{initial}\right) }$ and a final macrostate $\Theta
^{\left( \text{final}\right) }$. The trajectory can be equally well
traversed in both directions. Given the Fisher-Rao information metric, I can
apply standard methods of Riemannian differential geometry to study the
information-geometric structure of the manifold $\mathcal{M}_{S}$ underlying
the entropic dynamics. Connection coefficients $\Gamma _{\mu \nu }^{\rho }$,
Ricci tensor $R_{\mu \nu }$, Riemannian curvature tensor $R_{\mu \nu \rho
\sigma }$, sectional curvatures $\mathcal{K}_{\mathcal{M}_{S}}$, scalar
curvature $\mathcal{R}_{\mathcal{M}_{S}}$, Weyl anisotropy tensor $W_{\mu
\nu \rho \sigma }$, Killing fields $\xi ^{\mu }$ and Jacobi fields $J^{\mu }$
can be calculated in the usual way.

To characterize the chaotic behavior of complex entropic dynamical systems,
I are mainly concerned with the signs of the scalar and sectional curvatures
of $\mathcal{M}_{S}$, the asymptotic behavior of Jacobi fields $J^{\mu }$ on 
$\mathcal{M}_{S}$, the existence of Killing vectors $\xi ^{\mu }$ (or
existence of a non-vanishing Weyl anisotropy tensor, the anisotropy of the
manifold underlying system dynamics plays a significant role in the
mechanism of instability) and the asymptotic behavior of the
information-geometrodynamical entropy (IGE) $\mathcal{S}_{\mathcal{M}_{S}}$
(see (\ref{IGE})). It is crucial to observe that true chaos is identified by
the occurrence of two features: 1) strong dependence on initial conditions
and exponential divergence of the Jacobi vector field intensity, i.e., 
\textit{stretching} of dynamical trajectories; 2) compactness of the
configuration space manifold, i.e., \textit{folding} of dynamical
trajectories. The negativity of the Ricci scalar $\mathcal{R}_{\mathcal{M}%
_{S}}$,%
\begin{equation}
\mathcal{R}_{\mathcal{M}_{S}}=R_{\mu \nu \rho \sigma }g^{\mu \rho }g^{\nu
\sigma }=\sum_{\rho \neq \sigma }\mathcal{K}_{\mathcal{M}_{S}}\left( e_{\rho
}\text{, }e_{\sigma }\right) \text{,}
\end{equation}%
implies the existence of expanding directions in the configuration space
manifold $\mathcal{M}_{s}$. Indeed, since $\mathcal{R}_{\mathcal{M}_{S}}$ is
the sum of all sectional curvatures of planes spanned by pairs of
orthonormal basis elements $\left\{ e_{\rho }=\partial _{\Theta _{\rho
}}\right\} $, the negativity of the Ricci scalar is only a \textit{sufficient%
} (not necessary) condition for local instability of geodesic flow. For this
reason, the negativity of the scalar provides a \textit{strong }criterion of
local instability. Scenarios may arise where negative sectional curvatures
are present, but the positive ones could prevail in the sum so that the
Ricci scalar is non-negative despite the instability in the flow in those
directions. Consequently, the signs of $\mathcal{K}_{\mathcal{M}_{S}}$ are
of primary significance for the proper characterization of chaos.

A powerful mathematical tool to investigate the stability or instability of
a geodesic flow is the Jacobi-Levi-Civita equation (JLC equation) for
geodesic spread,%
\begin{equation}
\frac{D^{2}J^{\mu }}{D\tau ^{2}}+R_{\nu \rho \sigma }^{\mu }\frac{\partial
\Theta ^{\nu }}{\partial \tau }J^{\rho }\frac{\partial \Theta ^{\sigma }}{%
\partial \tau }=0\text{.}
\end{equation}%
The JLC-equation covariantly describes how nearby geodesics locally scatter
and relates the stability or instability of a geodesic flow with curvature
properties of the ambient manifold. Finally, the asymptotic regime of
diffusive evolution describing the possible exponential increase of average
volume elements on $\mathcal{M}_{s}$ provides another useful indicator of
dynamical chaoticity. The exponential instability characteristic of chaos
forces the system to rapidly explore large areas (volumes) of the
statistical manifold. It is interesting to note that this asymptotic
behavior appears also in the conventional description of quantum chaos where
the entropy \ (von Neumann) increases linearly at a rate determined by the
Lyapunov exponents. The linear increase of entropy as a quantum chaos
criterion was introduced by Zurek and Paz \cite{zurek1}. In my
information-geometric approach a relevant quantity that can be useful to
study the degree of instability characterizing ED models is the information
geometrodynamical entropy (IGE) defined as \cite{cafaro2},%
\begin{equation}
\mathcal{S}_{\mathcal{M}_{s}}\left( \tau \right) \overset{\text{def}}{=}%
\underset{\tau \rightarrow \infty }{\lim }\log \mathcal{V}_{\mathcal{M}_{s}}%
\text{ with }\mathcal{V}_{\mathcal{M}_{s}}\left( \tau \right) =\frac{1}{\tau 
}\dint\limits_{0}^{\tau }d\tau ^{\prime }\left( \underset{\mathcal{M}_{s}}{%
\int }\sqrt{g}d^{2l}\Theta \right)  \label{IGE}
\end{equation}%
and $g=\left\vert \det \left( g_{\mu \nu }\right) \right\vert $. IGE is the
asymptotic limit of the natural logarithm of the statistical weight defined
on $\mathcal{M}_{s}$ and represents a measure of temporal complexity of
chaotic dynamical systems whose dynamics is underlined by a curved
statistical manifold. In conventional approaches to chaos, the notion of
entropy is introduced, in both classical and quantum physics, as the missing
information about the systems fine-grained state \cite{caves}. For a
classical system, suppose that the phase space is partitioned into very
fine-grained cells of uniform volume $\Delta v$, labelled by an index $j$.
If one does not know which cell the system occupies, one assigns
probabilities $p_{j}$ to the various cells; equivalently, in the limit of
infinitesimal cells, one can use a phase-space density $\rho \left(
X_{j}\right) =\frac{p_{j}}{\Delta v}$. Then, in a classical chaotic
evolution, the asymptotic expression of the information needed to
characterize a particular coarse-grained trajectory out to time $\tau $ is
given by the Shannon information entropy (measured in bits),%
\begin{eqnarray}
\mathcal{S}_{\text{classical}}^{\left( \text{chaotic}\right) } &=&-\int
dX\rho \left( X\right) \log _{2}\left( \rho \left( X\right) \Delta v\right) 
\notag \\
&=&-\sum_{j}p_{j}\log _{2}p_{j}\sim \mathcal{K}\tau \text{.}
\end{eqnarray}%
where $\rho \left( X\right) $ is the phase-space density and $p_{j}=\frac{%
v_{j}}{\Delta v}$ is the probability for the corresponding coarse-grained
trajectory. $S_{\text{classical}}^{\left( \text{chaotic}\right) }$ is the
missing information about which fine-grained cell the system occupies. The
quantity $\mathcal{K}$ represents the linear rate of information increase
and it is called the Kolmogorov-Sinai entropy (or metric entropy) ($\mathcal{%
K}$ is the sum of positive Lyapunov exponents, $\mathcal{K}=\sum_{j}\lambda
_{j}$ ). $\mathcal{K}$ quantifies the degree of classical chaos.

\section{The information geometrodynamical approach to chaos: Applications}

In this Section I present three applications of the IGAC. First, I study the
chaotic behavior of an ED Gaussian model describing an arbitrary system of $l
$ degrees of freedom and show that the hyperbolicity of the non-maximally
symmetric $2l$-dimensional statistical manifold $\mathcal{M}_{s}$ underlying
such ED Gaussian model leads to linear information geometrodynamical entropy
(IGE) growth and to exponential divergence of the Jacobi vector field
intensity. I also present an information-geometric analogue of the Zurek-Paz
quantum chaos criterion of linear entropy growth. This analogy is presented
by studying the information geometrodynamics of ensemble of random frequency
macroscopic inverted harmonic oscillators. Finally, I apply the IGAC to
study the entropic dynamics on curved statistical manifolds induced by
classical probability distributions of common use in the study of regular
and chaotic quantum energy level statistics. In doing so, I suggest an
information-geometric characterization of regular and chaotic quantum energy
level statistics.

As I said in the Introduction, I have omitted technical details that will
appear elsewhere. However, my previous works (especially (\cite{cafaro2}))
may be very useful references in order to clarify the following applications.

\subsection{Chaotic behavior of an entropic dynamical Gaussian model}

As a first example, I apply my IGAC to study the dynamics of a system with $%
l $ degrees of freedom, each one described by two pieces of relevant
information, its mean expected value and its variance (Gaussian statistical
macrostates). The line element $ds^{2}=g_{\mu \nu }\left( \Theta \right)
d\Theta ^{\mu }d\Theta ^{\nu }$ on $\mathcal{M}_{s}$ is defined by,%
\begin{equation}
ds^{2}=\dsum\limits_{k=1}^{l}\left( \frac{1}{\sigma _{k}^{2}}d\mu _{k}^{2}+%
\frac{2}{\sigma _{k}^{2}}d\sigma _{k}^{2}\right) \text{, with }\mu \text{, }%
\nu =1\text{,..., }2l\text{.}
\end{equation}%
This leads to consider an ED model on a non-maximally symmetric $2l$%
-dimensional statistical manifold $\mathcal{M}_{s}$. It is shown that $%
\mathcal{M}_{s}$ possesses a constant negative Ricci curvature that is
proportional to the number of degrees of freedom of the system, $R_{\mathcal{%
M}_{s}}=-l$. It is shown that the system explores statistical volume
elements on $\mathcal{M}_{s}$ at an exponential rate. The information
geometrodynamical entropy $\mathcal{S}_{\mathcal{M}_{s}}$\ increases
linearly in time (statistical evolution parameter) and is moreover,
proportional to the number of degrees of freedom of the system, $\mathcal{S}%
_{\mathcal{M}_{s}}$ $\overset{\tau \rightarrow \infty }{\sim }l\lambda \tau $%
. The parameter $\lambda $ characterizes the family of probability
distributions on $\mathcal{M}_{s}$. The asymptotic linear
information-geometrodynamical entropy growth may be considered the
information-geometric analogue of the von Neumann entropy growth introduced
by Zurek-Paz, a \textit{quantum} feature of chaos. The geodesics on $%
\mathcal{M}_{s}$ are hyperbolic trajectories. Using the Jacobi-Levi-Civita
(JLC) equation for geodesic spread, I show that the Jacobi vector field
intensity $J_{\mathcal{M}_{s}}$ diverges exponentially and is proportional
to the number of degrees of freedom of the system, $J_{\mathcal{M}_{s}}$ $%
\overset{\tau \rightarrow \infty }{\sim }l\exp \left( \lambda \tau \right) $%
. The exponential divergence of the Jacobi vector field intensity $J_{%
\mathcal{M}_{s}}$ is a \textit{classical} feature of chaos. Therefore, we
conclude \ that 
\begin{equation}
\mathcal{R}_{\mathcal{M}_{s}}=-l\text{, }J_{\mathcal{M}_{s}}\overset{\tau
\rightarrow \infty }{\sim }l\exp \left( \lambda \tau \right) \text{, }%
\mathcal{S}_{\mathcal{M}_{s}}\overset{\tau \rightarrow \infty }{\sim }%
l\lambda \tau \text{.}
\end{equation}%
Thus, $\mathcal{R}_{\mathcal{M}_{s}}$, $\mathcal{S}_{\mathcal{M}_{s}}$ and $%
J_{\mathcal{M}_{s}}$ behave as proper indicators of chaoticity and are
proportional to the number of Gaussian-distributed microstates of the
system. This proportionality, even though proven in a very special case,
leads to conclude there may be a substantial link among these
information-geometric indicators of chaoticity.

\subsection{Ensemble of random frequency macroscopic inverted harmonic
oscillators}

In our second example, I employ ED and "Newtonian Entropic Dynamics" (NED) 
\cite{cafaro4}. In this NED, we explore the possibility of using well
established principles of inference to derive Newtonian dynamics from
relevant prior information codified into an appropriate statistical
manifold. The basic assumption is that there is an irreducible uncertainty
in the location of particles so that the state of a particle is defined by a
probability distribution. The corresponding configuration space is a
statistical manifold $\mathcal{M}_{s}$ the geometry of which is defined by
the Fisher-Rao information metric. The trajectory follows from a principle
of inference, the method of Maximum Entropy. There is no need for additional
"physical" postulates such as an action principle or equation of motion, nor
for the concept of mass, momentum and of phase space, not even the notion of
time. The resulting "entropic" dynamics reproduces Newton's mechanics for
any number of particles interacting among themselves and with external
fields. Both the mass of the particles and their interactions are explained
as a consequence of the underlying statistical manifold.

In my special application, I consider a manifold with a line element $%
ds^{2}=g_{\mu \nu }\left( \Theta \right) d\Theta ^{\mu }d\Theta ^{\nu }$
(with $\mu $, $\nu =1$,..., $l$) given by,%
\begin{equation}
ds^{2}=\left[ 1-\Phi \left( \Theta \right) \right] \delta _{\mu \nu }\left(
\Theta \right) d\Theta ^{\mu }d\Theta ^{\nu }\text{, }\Phi \left( \Theta
\right) =\overset{l}{\underset{k=1}{\sum }}u_{k}\left( \theta _{k}\right) 
\text{ }
\end{equation}%
where%
\begin{equation}
u_{k}\left( \theta _{k}\right) =-\frac{1}{2}\omega _{k}^{2}\theta _{k}^{2}%
\text{, }\theta _{k}=\theta _{k}\left( s\right) \text{.}
\end{equation}%
The geodesic equations for the macrovariables $\theta _{k}\left( s\right) $
are strongly \textit{nonlinear }and their integration is not trivial.
However, upon a suitable change of the affine parameter $s$ used in the
geodesic equations, I may simplify the differential equations for the
macroscopic variables parametrizing points on the manifold $\mathcal{M}_{s}$
with metric tensor $g_{\mu \nu }$. Recalling that the notion of chaos is
observer-dependent and upon changing the affine parameter from $s$ to $\tau $
in such a way that $ds^{2}=2\left( 1-\Phi \right) ^{2}d\tau ^{2}$, I obtain
new geodesic equations describing a set of macroscopic inverted harmonic
oscillators (IHOs). In order to ensure the compactification of the parameter
space of the system (and therefore $\mathcal{M}_{s}$ itself), we choose a
Gaussian distributed frequency spectrum for the IHOs. Thus, with this choice
of frequency spectrum, the folding mechanism required for true chaos is
restored in a statistical (averaging over $\omega $ and $\tau $) sense. Upon
integrating these differential equations, I obtain the expression for the
asymptotic behavior of the IGE $\mathcal{S}_{\mathcal{M}_{s}}$, namely%
\begin{equation}
\mathcal{S}_{\mathcal{M}_{s}}\left( \tau \right) \overset{\tau \rightarrow
\infty }{\sim }\Lambda \tau \text{, }\Lambda =\overset{l}{\underset{i=1}{%
\sum }}\omega _{i}\text{.}
\end{equation}%
This result may be considered the information-geometric analogue of the
Zurek-Paz model used to investigate the implications of decoherence for
quantum chaos. In their work, Zurek and Paz considered a chaotic system, a
single unstable harmonic oscillator characterized by a potential $V\left(
x\right) =-\frac{\Omega ^{2}x^{2}}{2}$ ($\Omega $ is the Lyapunov exponent),
coupled to an external environment. In the \textit{reversible classical
limit }\cite{zurek2}, the von Neumann entropy of such a system increases
linearly at a rate determined by the Lyapunov exponent, 
\begin{equation}
\mathcal{S}_{\text{quantum}}^{\left( \text{chaotic}\right) }\left( \tau
\right) \overset{\tau \rightarrow \infty }{\sim }\Omega \tau \text{,}
\end{equation}%
with $\Omega $ playing the role of the Lyapunov exponent.

\subsection{Information geometrodynamics of regular and chaotic quantum spin
chains}

In my final example, I use my IGAC to study the entropic dynamics on curved
statistical manifolds induced by classical probability distributions of
common use in the study of regular and chaotic quantum energy level
statistics. Recall that the theory of quantum chaos (quantum mechanics of
systems whose classical dynamics are chaotic) is not primarily related to
few-body physics. Indeed, in real physical systems such as many-electron
atoms and heavy nuclei, the origin of complex behavior is the very strong
interaction among many particles. To deal with such systems, a famous
statistical approach has been developed which is based upon the Random
Matrix Theory (RMT). The main idea of this approach is to neglect the
detailed description of the motion and to treat these systems statistically
bearing in mind that the interaction among particles is so complex and
strong that generic properties are expected to emerge. Once again, this is
exactly the philosophy underlining the ED approach to complex dynamics. It
is known \cite{prosen} that integrable and chaotic quantum antiferromagnetic
Ising chains are characterized by asymptotic logarithmic and linear growths
of their operator space entanglement entropies, respectively. In this last
example, I consider the information-geometrodynamics of a Poisson
distribution coupled to an Exponential bath (spin chain in a \textit{%
transverse} magnetic field, regular case) and that of a Wigner-Dyson
distribution coupled to a Gaussian bath (spin chain in a \textit{tilted}
magnetic field, chaotic case). Remarkably, I show that in the former case
the IGE exhibits asymptotic logarithmic growth while in the latter case the
IGE exhibits asymptotic linear growth. In the regular case, the line element 
$ds_{\text{integrable}}^{2}=$ $ds_{\text{Poisson}}^{2}+ds_{\text{Exponential}%
}^{2}$ on the statistical manifold $\mathcal{M}_{s}$ is given by%
\begin{equation}
ds_{\text{integrable}}^{2}=\frac{1}{\mu _{A}^{2}}d\mu _{A}^{2}+\frac{1}{\mu
_{B}^{2}}d\mu _{B}^{2}
\end{equation}%
where the macrovariable $\mu _{A}$ is the average spacing of the energy
levels and $\mu _{B}$ is the average intensity of the magnetic energy
arising from the interaction of the \textit{transverse} magnetic field with
the spin $\frac{1}{2}$\ particle magnetic moment. In such a case, I show
that the asymptotic behavior of $\ \mathcal{S}_{\mathcal{M}_{s}}^{\left( 
\text{integrable}\right) }$ is sub-linear in $\tau $ (logarithmic IGE
growth), 
\begin{equation}
\mathcal{S}_{\mathcal{M}_{s}}^{\left( \text{integrable}\right) }\left( \tau
\right) \overset{\tau \rightarrow \infty }{\sim }\log \tau \text{.}
\end{equation}%
Finally, in the chaotic case, the line element $ds_{\text{chaotic}}^{2}=ds_{%
\text{Wigner-Dyson}}^{2}+ds_{\text{Gaussian}}^{2}$ on the statistical
manifold $\mathcal{M}_{s}$ is given by%
\begin{equation}
ds_{\text{chaotic}}^{2}=\frac{4}{\mu _{A}^{\prime 2}}d\mu _{A}^{\prime 2}+%
\frac{1}{\sigma _{B}^{\prime 2}}d\mu _{B}^{\prime 2}+\frac{2}{\sigma
_{B}^{\prime 2}}d\sigma _{B}^{\prime 2}
\end{equation}%
where the (nonvanishing) macrovariable $\mu _{A}^{\prime }$ is the average
spacing of the energy levels, $\mu _{B\text{ }}^{\prime }$and $\sigma
_{B}^{\prime }$ are the average intensity and variance, respectively of the
magnetic energy arising from the interaction of the \textit{tilted} magnetic
field with the spin $\frac{1}{2}$\ particle magnetic moment. In this case, I
show that asymptotic behavior of $\ \mathcal{S}_{\mathcal{M}_{s}}^{\left( 
\text{chaotic}\right) }$ is linear in $\tau $ (linear IGE growth), 
\begin{equation}
\mathcal{S}_{\mathcal{M}_{s}}^{\left( \text{chaotic}\right) }\left( \tau
\right) \overset{\tau \rightarrow \infty }{\sim }\tau \text{.}
\end{equation}%
The equations for $\mathcal{S}_{\mathcal{M}_{s}}^{\left( \text{integrable}%
\right) }$ and $\mathcal{S}_{\mathcal{M}_{s}}^{\left( \text{chaotic}\right)
} $ are the information-geometric analogue of the entanglement entropies
defined in standard quantum information theory in the regular and chaotic
cases, respectively. In addition, I emphasize that the statistical volume
element $\mathcal{V}_{\mathcal{M}_{s}}\left( \tau \right) $ (see (\ref{IGE}%
)) may play a similar role as the computational complexity in conventional
quantum information theory. These results warrant deeper analysis to be
fully understood.

\section{Conclusion}

In this paper I proposed a theoretical information-geometric framework
suitable to characterize chaotic dynamical behavior of arbitrary complex
systems on curved statistical manifolds. Specifically, an
information-geometric analogue of the Zurek-Paz quantum chaos criterion of
linear entropy growth and an information-geometric characterization of
regular and chaotic quantum energy level statistics was presented. I hope
that my work convincingly shows that this information-geometric approach may
be useful in providing a unifying criterion of chaos of both classical and
quantum varieties, thus deserving further research and developments.

The descriptions of a classical chaotic system of arbitrary interacting
degrees of freedom, deviations from Gaussianity and chaoticity arising from
fluctuations of positively curved statistical manifolds are currently under
investigation. I am also investigating the possibility to extend the IGAC to
quantum Hilbert spaces constructed from classical curved statistical
manifolds and I am considering the information-geometric macroscopic
versions of the Henon-Heiles and Fermi-Pasta-Ulam $\beta $-models to study
chaotic geodesic flows on statistical manifolds.

\begin{acknowledgments}
I am grateful to Sean Alan Ali, Ariel Caticha and Adom Giffin for very
useful comments, discussions and for their previous collaborations. I extend
thanks to Cedric Beny, Michael Frey and Jeroen Wouters for their interest
and/or useful comments on my research during the NIC@QS07 in Erice, Ettore
Majorana Centre.
\end{acknowledgments}

\end{document}